\documentclass[sigconf]{acmart}
\AtBeginDocument{%
  }

\setcopyright{acmlicensed}
\copyrightyear{2018}
\acmYear{2018}
\acmDOI{XXXXXXX.XXXXXXX}
\acmConference[Conference acronym 'XX]{Make sure to enter the correct
  conference title from your rights confirmation email}{June 03--05,
  2018}{Woodstock, NY}
\acmISBN{978-1-4503-XXXX-X/2018/06}

\usepackage{multirow}

\copyrightyear{2025}
\acmYear{2025}
\setcopyright{acmlicensed}\acmConference[SIGIR '25]{Proceedings of the 48th International ACM SIGIR Conference on Research and Development in Information Retrieval}{July 13--18, 2025}{Padua, Italy}
\acmBooktitle{Proceedings of the 48th International ACM SIGIR Conference on Research and Development in Information Retrieval (SIGIR '25), July 13--18, 2025, Padua, Italy}
\acmDOI{10.1145/3726302.3730215}
\acmISBN{979-8-4007-1592-1/2025/07}

\begin{document}

\title{Interest Changes: Considering User Interest Life Cycle in Recommendation System}


\author{Yinjiang Cai}
\email{caiyinjiang@corp.netease.com}
\authornote{Corresponding author}
\affiliation{%
  \institution{Hangzhou NetEase Entertainment Technology Co., Ltd}
  \city{Hangzhou}
  \country{China}
}

\author{Jiangpan Hou}
\email{houjiangpan@corp.netease.com}
\affiliation{%
  \institution{Hangzhou NetEase Entertainment Technology Co., Ltd}
  \city{Hangzhou}
  \country{China}
}

\author{Yangping Zhu}
\email{hzzhuyangping@corp.netease.com}
\affiliation{%
  \institution{Hangzhou NetEase Entertainment Technology Co., Ltd}
  \city{Hangzhou}
  \country{China}
}

\author{Yuan Nie}
\email{hznieyuan@corp.netease.com}
\affiliation{%
  \institution{Hangzhou NetEase Entertainment Technology Co., Ltd}
  \city{Hangzhou}
  \country{China}
}

\renewcommand{\shortauthors}{Yinjiang Cai, Jiangpan Hou, Yangping Zhu, \& Yuan Nie}

\begin{abstract}
  In recommendation systems, user interests are always in a state of constant flux. Typically, a user interest experiences a emergent phase, a stable phase, and a declining phase, which are referred to as the "user interest life-cycle".
  Recent papers on user interest modeling have primarily focused on how to compute the correlation between the target item and user's historical behaviors, without thoroughly considering the life-cycle features of user interest. In this paper, we propose an effective method called \textbf{D}eep \textbf{I}nterest \textbf{L}ife-cycle \textbf{N}etwork (DILN), which not only captures the interest life-cycle features efficiently, but can also be easily integrated to existing ranking models. DILN contains two key components: \textbf{Interest Life-cycle Encoder Module} constructs historical activity histograms of the user interest and then encodes them into dense representation. \textbf{Interest Life-cycle Fusion Module} injects the encoded dense representation into multiple expert networks, with the aim of enabling the specific phase of interest life-cycle to activate distinct experts. Online A/B testing reveals that DILN achieves significant improvements of +0.38\% in CTR, +1.04\% in CVR and +0.25\% in duration per user, which demonstrates its effectiveness. In addition, DILN inherently increase the exposure of users' emergent and stable interests while decreasing the exposure of declining interests. DILN has been deployed on the Lofter App. 

\end{abstract}

\ccsdesc[500]{Information systems~Recommendation systems}

\keywords{Recommendation System, Interest Modeling, Interest Life-Cycle}


\maketitle

\section{Introduction}
User interest modeling is one of the important research directions in the field of recommendation systems. 
Recent research has primarily focused on modeling the diversity\cite{mind, comirec} and precision\cite{din, dien, sim} of user interests. However, critical characteristics of user interests, their \textbf{interest life-cycles} have often been overlooked. User interests typically undergo multiple phases: emergent phase, stable phase, and declining phase. (1) During the \textbf{emergent phase}, users exhibit strong curiosity and engagement, with the frequency and intensity of activities related to the interest gradually increasing. However, due to the sparse behavioral signals in the emergent phase, existing methods fail to promptly model and promote the emergent interest. (2) In the \textbf{stable phase}, both the frequency and intensity of user activities stabilize at a high level. Current models primarily improve modeling accuracy through lifelong sequential search strategies\cite{sim, eta, sdim, twin}. (3) In the \textbf{declining phase}, users' attention toward the interest gradually diminishes. Due to the abundant historical behaviors, existing ranking model tend to dispatch such content based on behavioral correlations, while ignoring the declining trend of the interest.

\begin{figure}[h]
  \centering
  \includegraphics[width=\linewidth]{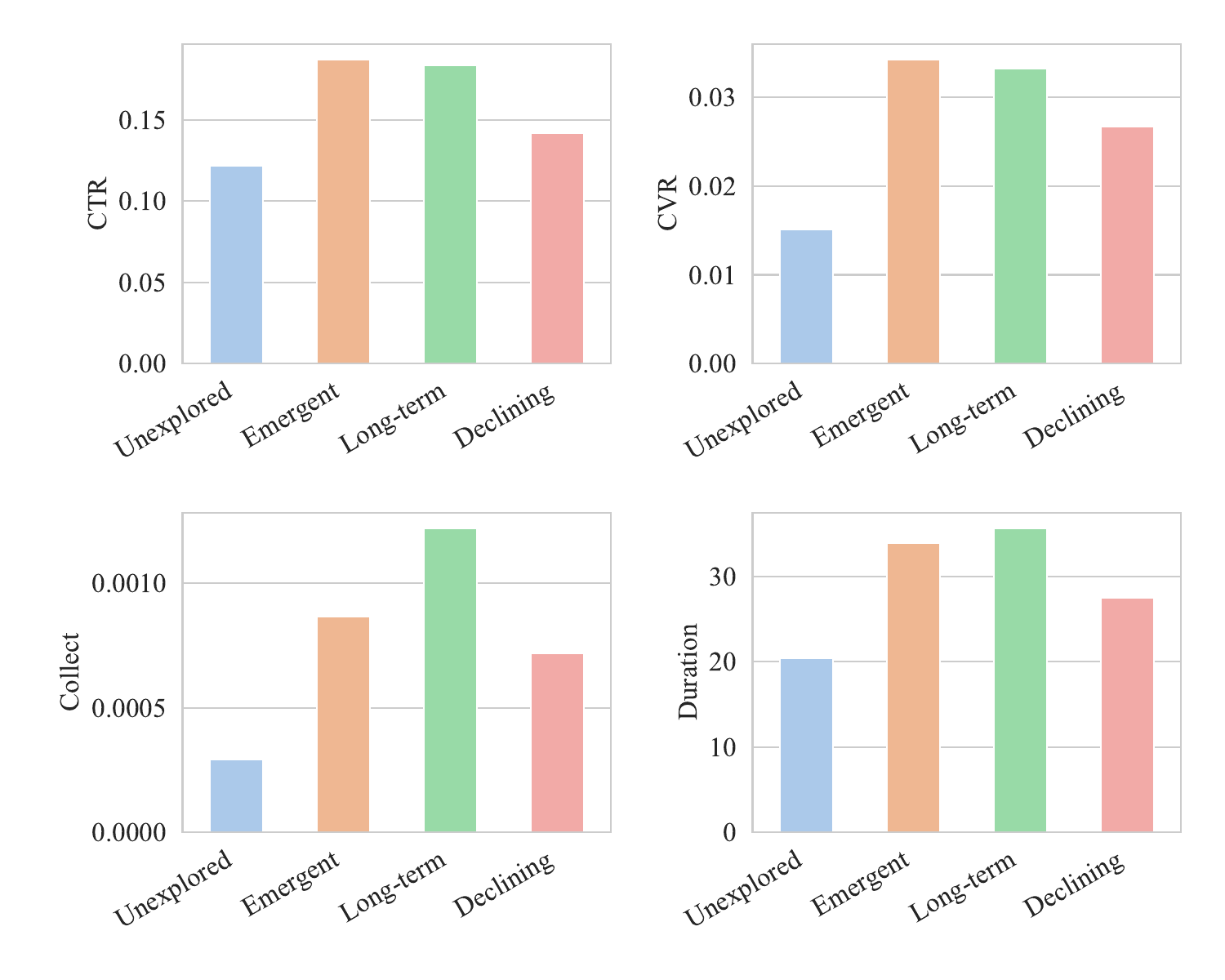}
  \caption{
    Comparison between different interest life-cycles. The y-axis represents task-specific efficiency for each interest category.
  }
\label{abstract_fig}
\end{figure}

Beyond the aforementioned differences, the efficiency of different interest life-cycles also exhibits significant variations across various tasks, providing valuable insights for further optimizing recommendation systems. In Figure \ref{abstract_fig}, we categorize recommendation results from the Lofter app into four classes based on  business logic: Unexplored, Emergent, Long-term and Declining interests. It is not difficult to notice that despite having far fewer historical behaviors than long-term interests, emergent interests still exhibit the highest efficiency across two tasks. On the other hand, while declining interests have rich historical behaviors, their overall efficiency is reduced due to diminishing relevance.
.

In this paper, we address the following challenges: (1) How to integrate the above life-cycle features of user interests into the recommendation systems? (2) How to utilize interest life-cycle features to improve the accuracy of user interest modeling? 
To address the first challenge, we propose a novel approach called Interest Life-cycle Encoder Module (ILEM), which constructs and encodes interest life-cycle features. By applying ILEM, interest life-cycle features are encoded as dense vectors, enabling seamless integration into any component of existing ranking models.

To address the second challenge, we propose the Interest Life-cycle Fusion Module (ILFM), which integrates the interest life-cycle features into multi-task recommendation systems. ILFM dynamically generates rescaling factors for the input features and representations of the hidden layers based on the current life-cycle of the user interest, applying the application of the Hadamard product to refine the results. In this way, ILFM can selectively emphasize critical features for each interest life-cycle and perform personalized transformations on the representations between hidden layers to optimize different recommendation tasks.

The key contributions of this work are summarized as follows:
\begin{itemize}
\item  We design the Interest Life-cycle Encoder Module to characterize the life-cycle features of user interests. To the best of our knowledge, we are the first to introduce interest life-cycle features into recommendation systems.
\item  We propose the Interest Life-cycle Fusion Module as an effective and universal block for integrating interest life-cycle features into the ranking model of recommendation systems.
\item  By conducting the offline and online experiments, we demonstrate the effectiveness of DILN and deploy it on the Lofter app.
\end{itemize}

\section{Related Work}
Modeling user interests is a key focus in industrial research, particularly in studying the evolution of user interests.
These approaches not only characterize the interests of users at present but also capture historical changes in user interests \cite{interest_trend}. 
DIEN \cite{dien} and DSIN \cite{dsin} utilize Recurrent Neural Networks (RNN) to model how user interests evolve over time. SIM \cite{sim} first introduced a search-based unit to model lifelong user interest sequences. 
Furthermore, several methods\cite{long_short_interest, long_short_interest_2, long_short_interest_3} proposed to fuse long-term and short-term interests to cover the evolution of the interest.

\section{Method}


\subsection{Interest Life-cycle Encoder Module}\label{subsec: fea_encoder}
\subsubsection{Feature engineering}
To describe the recent activity trends of users over candidate interests, we first apply a General Search Unit (GSU) to retrieve the most relevant user behaviors $B^t = [b^t_1; b^t_2; ..., b^t_N]$ with respect to the candidate item, where $t$ denotes the action type (including exposure, click and interaction). 
For simplicity, we restrict the search range to the recent $K$ activities and the number of search results to $N$. 

\begin{figure}[h]
  \centering
  \vspace{-5pt} 
  \includegraphics[width=\linewidth]{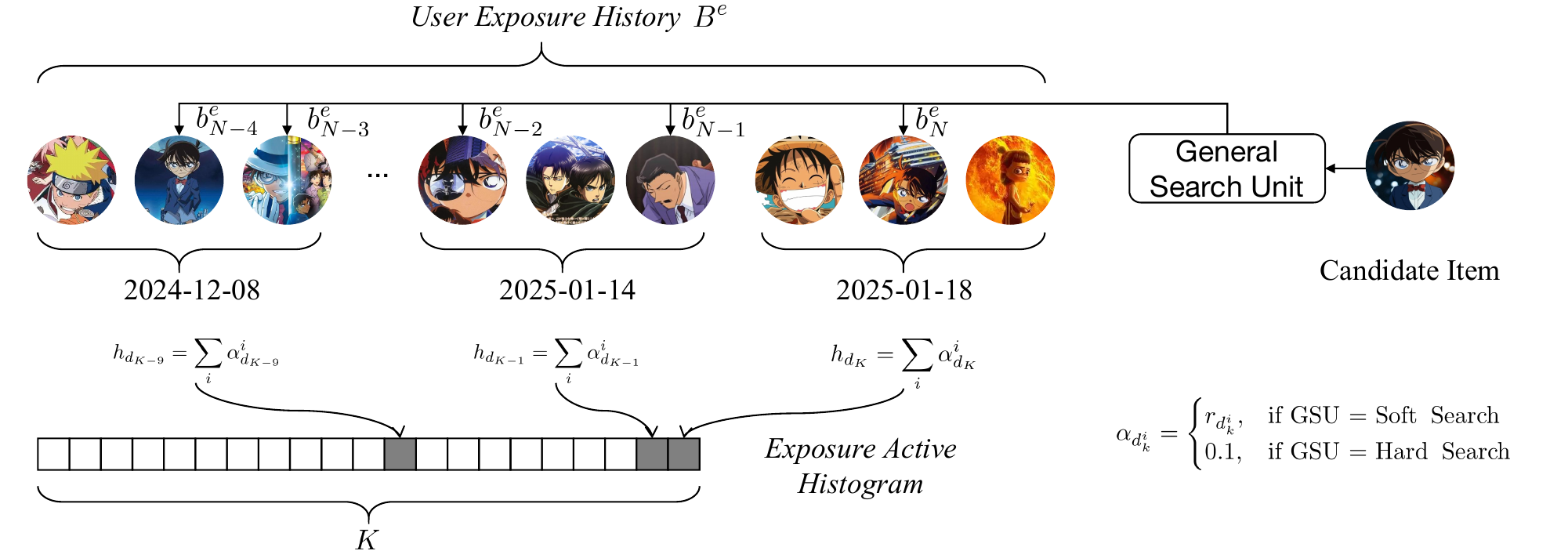}
  \caption{
    Illustration of feature engineering applied to user exposure history. 
  }
  \vspace{-5pt} 
\label{feature_fig}
\end{figure}

In Figure \ref{feature_fig}, we take the action type "exposure" as an example. For each user active date containing search results, the sum of the relevance scores $\alpha_{d_i}$ of items under that date forms a data point $h_{d_i}$ in the histogram.
To accommodate different types of GSU, we use the search relevance score $r_{d_k^i}$ as $\alpha_{d_k^i}$ for Soft Search, while setting $\alpha_{d_k^i}$ to 0.1 for Hard Search.

Ultimately, this vector of length K captures the distribution of user activity intensity for the candidate interest over recent K activities. Using this approach, we can derive the interest life-cycle histograms across various behaviors including exposure, click, and interaction.

\subsubsection{Histogram Encoder}\label{subsec: encoder}
The aforementioned histograms inherently possess temporal characteristics. To model these temporal features effectively, we employ a multi-layer convolutional neural network. 
Specifically, we utilize convolutional kernels with kernel sizes of 5, 3, 2, and 8, 16, 32 filters respectively, to perform 1D convolutional operation. Finally, a linear layer transforms the convolved features into fixed-dimension vectors.

Numerous methods exist for time series modeling, such as CNN\cite{tcn, wavenet} and Transformer\cite{lim2021temporal, zhou2021informer}. Here, we choose CNN due to its balance of time complexity and efficiency. While more complex and effective approaches exist, they are beyond the scope of this paper.

\begin{figure*}[t]
\centering
\includegraphics[width=1.0\textwidth]{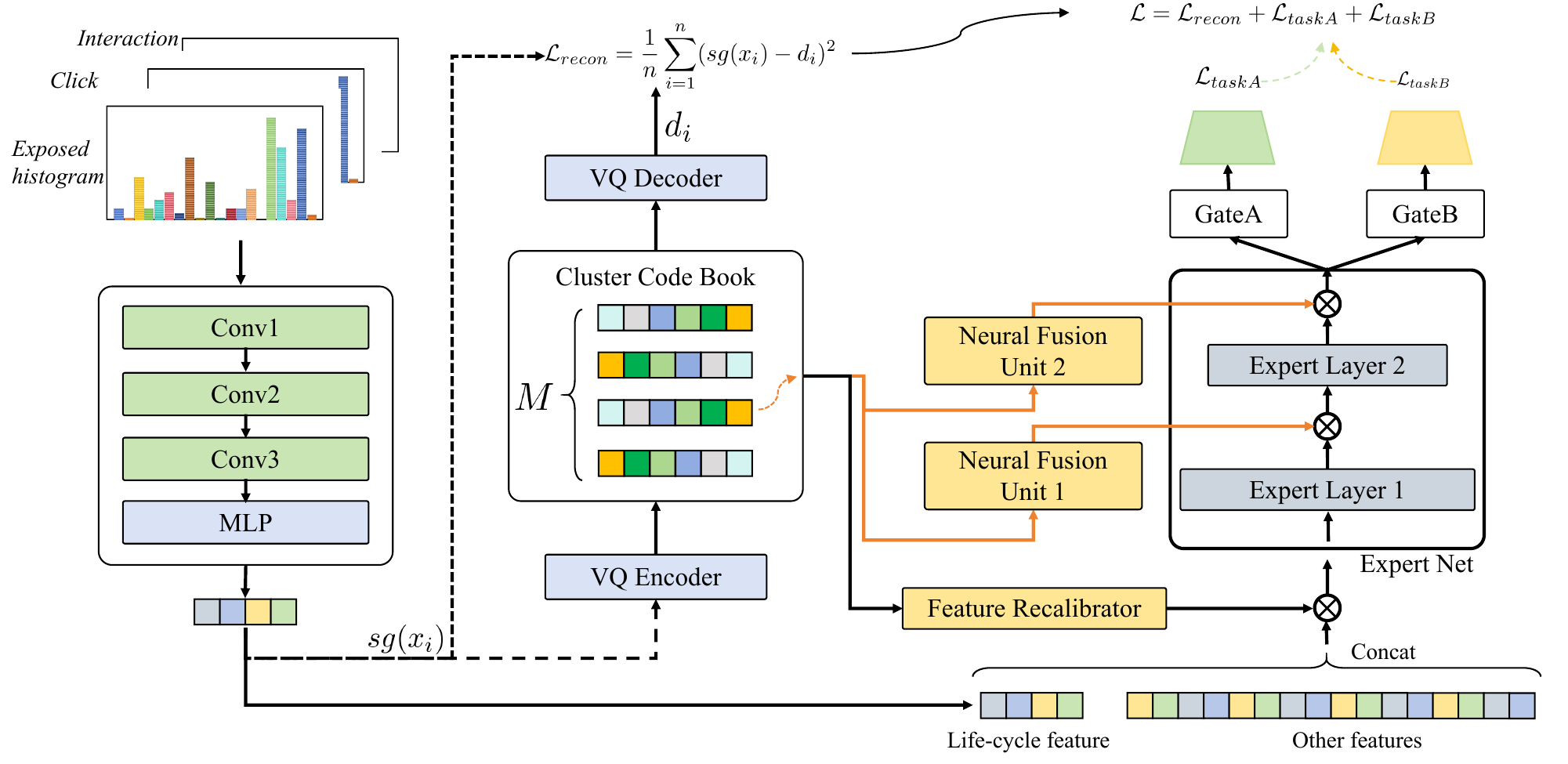} 
\caption{The architecture of the proposed Deep Interest Life-Cycle Network.}
\label{modelStructure}
\end{figure*}

\subsubsection{Life-Cycle VQ Cluster}\label{subsec: vq}
In this part, we employ Vector Quantization (VQ) technique\cite{vqvae} to dynamically divide all samples into $M$ clusters based on the above interest life-cycle vector, with each cluster representing a distinct interest life cycle. The VQ cluster contains three parts: VQ encoder, VQ search and VQ decoder, as defined in Equation \ref{vq}:

\begin{equation}
    \begin{aligned}
\label{vq}
    xc_i = enco&der(sg(x_i)) ~\\
      c_i = NearestCod&e(xc_i, Q),  Q \in \mathbb{R}^{M \times d} ~\\
      d_i = decoder(xc_i &+ sg(c_i-xc_i)) ~\\
      \mathcal L _{recon} = \frac{1}{n} \sum_{i=1}^{n} &(sg(x_i) - d_i)^2
    \end{aligned}
\end{equation}

where $x_i$ denotes the output of histogram encoder, $sg$ represents the stop-gradient operation applied to the tensor. In VQ encoder, each feature $x_i$ is compressed by multi-layer perceptrons. By searching the VQ codebook, we identify the nearest cluster center $c_i$ to the compressed vector $xc_i$. To supervise the learning of VQ cluster effectively, we introduce a decoder to reconstruct the raw feature $x_i$ and employ Mean Square Loss (MSE) to supervise the distance between the decoder output $d_i$ and the $x_i$. 
Ultimately, the Life-Cycle VQ Cluster block provides the cluster center $c_i$ derived from search, as a dense representation of the life-cycle for the current candidate interest.

\subsection{Interest Life-cycle Fusion Module}\label{subsec: ilfm}
As illustrated in Figure \ref{abstract_fig}, interests at different life-cycles exhibit significant variations across different recommendation tasks. To empower ranking models to better recognize and address these variations, we propose an Interest Life-cycle Fusion Module to explicitly inject interest life-cycle features into ranking models. ILFM is an effective and versatile module that can be seamlessly integrated into existing ranking models. In this work, we construct the ILFM based on Multi-gate Mixture-of-Experts (MMOE)\cite{mmoe}, which is a widely adopted multi-task framework.

\subsubsection{Feature Recalibrator}
For samples within different interest life-cycles, the importance of each feature varies significantly. 
For instance, features such as historical activity and lifelong sequence should be downweighted for declining interests to prevent their over-recommendation. Conversely, features related to short-term activity require enhancement to better capture emergent interests.
Here, we propose a feature recalibrator to dynamically re-weight each feature based on the interest life-cycles of the samples.
The feature recalibrator contains two layers, which are formulated as follows:

\begin{equation}
    \begin{aligned}
      g_i = Relu&(\mathbf{W}c_i + \mathbf{b}) ~\\
      g'_i  = \gamma * Sigm&oid(\mathbf{W'}g_i + \mathbf{b'}) ~\\
      z_i' = g'_i &\odot z_i ~\\
    \end{aligned}
\end{equation}

where $z_i$ denotes the share-bottom input and $c_i$ denotes the dense representation of interest life-cycle derived from the above interest life-cycle encoder. $\mathbf{W}$, $\mathbf{b}$, $\mathbf{W'}$, and $\mathbf{b'}$ are learnable parameters.  $\gamma$ is the scaling factor which is set as 2, and $\odot$ denotes element-wise multiplication.

\subsubsection{Neural Fusion Unit}

As is widely recognized, the expert nets in MMOE are responsible for task-specific encoding based on input features. Given the divergent performance of interest life-cycles across different tasks, we explicitly conduct hierarchical interactions between the interest life-cycle features and the outputs of each layer within the expert nets.

\begin{equation}
    \begin{aligned}
        f_i = \gamma  * Sigm&oid(\mathbf{W_g}c_i+b_g) ~\\
        x_e = f_i \odot \sigma&(\mathbf{W_e}x+b_e)
    \end{aligned}
\end{equation}

where $\mathbf{W_e}$ and $\mathbf{b_e}$ denote parameters in each expert net. $\mathbf{W_g}$ and $\mathbf{b_g}$ are learnable parameters to generate the scaling factor based on the cluster of interest life-cycles.

\section{Experiment}
In this section, we conduct comprehensive offline and online experiments with the aim of answering the following questions.

\begin{itemize}
\item \textbf{RQ1}: How does DILN perform on large-scale industrial datasets compared to existing methods?

\item \textbf{RQ2}: Can the proposed method DILN achieve performance improvements in online recommendation systems?

\item \textbf{RQ3}: What is the impact of incorporating interest life-cycle features in the real-world recommendation systems?
\end{itemize}

\subsection{Experimental Settings}
We evaluate DILN on two datasets:
\textbf{KuaiRand} \cite{gao2022kuairand} is a public recommendation dataset containing real user interactions, including over 27,000 users and 300 million interactions. Since the dataset contains only 30 days of data, we use the first 20 days for feature construction, the next 8 days for training, the following 1 day as the validation set, and the remaining 1 day as test set. \textbf{Industrial Dataset}: This dataset was collected from Lofter App with millions of daily active users. We gathered exposure logs over a 10-day period, including more than 6 million users and 2.7 billion samples. We use the first 8 days as the training set,  the following 1 day as the validation set, the last day as the test set. For both datasets, we set the number of GSU results $N$ to 100, set the length of histograms $K$ to 20, and set the number of VQ clusters $M$ to 10.

\begin{table}[t]
\caption{Model Comparison with GAUC on offline datasets}
\label{table3}
\centering
\begin{tabular}{c r r r r}
\toprule
\multirow{2}{*}{Method} & \multicolumn{2}{c}{KuaiRand} & \multicolumn{2}{c}{Industry}\\
\cmidrule(lr){2-3}
\cmidrule(lr){4-5}
&CTR& \multicolumn{1}{c}{CVR} &CTR& \multicolumn{1}{c}{CVR}\\

\midrule

SIM         & 0.6436    & 0.6364    & 0.5832   & 0.6636 \\

SIM+ILEM     & 0.6708    & 0.6497    & 0.5921   & 0.6726 \\

DILN        & \textbf{0.6726}    & \textbf{0.6512}    & \textbf{0.5934}   & \textbf{0.6751} \\

\bottomrule
\end{tabular}
\end{table}

\subsection{Offline Result}
To comprehensively evaluate the effectiveness of DILN and its components, we use MMOE combined with SIM (Hard Search) as the baseline. We use the GAUC as an offline metric. As shown in Table $1$, by simply adding the ILEM block on the baseline model, both the CTR and CVR task achieve significant improvement. When applying the ILFM block additionally, two tasks achieve further improvements, as the ILFM explicitly injects interest life-cycle features into the expert nets, thereby enabling them to better capture patterns between interest life-cycle and specific task objectives.

To better illustrate the role of the VQ Cluster block, we visualize the activated VQ clusters of samples based on pre-defined interest categories. As shown in Figure 4, the distribution of activated clusters undergoes substantial changes as the interest life-cycle evolves. For example, the Unexplored interests predominantly activate cluster 0, while the emergent interests primarily activate clusters 3 and 4. 
This demonstrates that our proposed VQ clustering module can effectively stratify the samples across different interest life-cycles.

\begin{figure}[h]
  \centering
  \includegraphics[width=\linewidth]{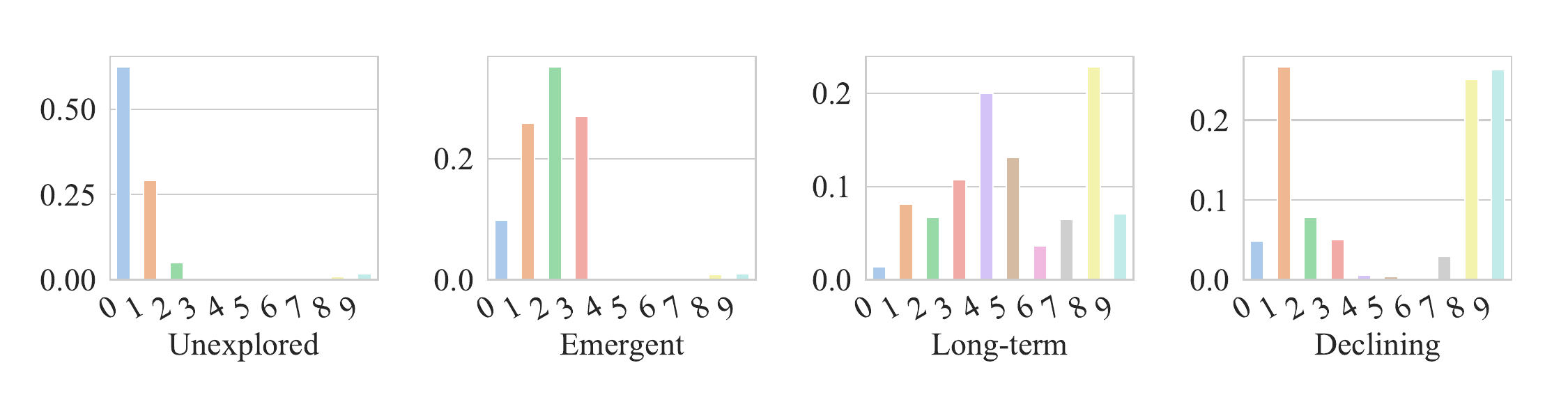}
  \caption{
    Illustration of activated VQ clusters across different interest life-cycles. The y-axis represents the activation probability of each cluster.
  }
\label{cluster_result}
\end{figure}

\subsection{Online Result}
To evaluate the real online improvements brought by DILN, we conducted online A/B testing on the Lofter platform for more than three weeks and involved over 20\% of users in the experiment group. 
We primarily observed three metrics: CTR, CVR, and Duration. 
The experimental group demonstrated that DILN achieved a 0.38\% improvement in CTR, a 1.04\% improvement in CVR and a 0.25\% improvement in duration, compared to our latest production model.

To address the above RQ 3, we analyzed the changes in the distribution across different stages of interest life cycles during the online experiment. From Table \ref{online}, we observed that DILN significantly reduced the impression of inefficient interests (Unexplored and Declining) while encouraging the recommendations of positive interests such as Emergent interest. Through interest life-cycle modeling, DILN improved the efficiency of dispatch across all types of interests.

\begin{table}[t]
\caption{Online dispatch differences between DILN and the baseline model.}
\label{online}
\centering
\begin{tabular}{c r r r r}
\toprule
            & Unexplored   & Emergent    & Long-term   &Declining \\

\midrule

Impression      & -2.17\%    & +5.11\%    & +0.0\%   & -4.4\%\\

CTR             & +0.15\%    & +0.60\%    & +0.4\%   & +1.15\%\\

CVR             & +1.33\%    & +0.1\%    & +1.16\%   & +2.20\%\\

\bottomrule
\end{tabular}
\end{table}

\section{Conclusion}
In this paper, we propose the DILN method to model interest life-cycle features in recommendation systems, a critical aspect that has often been overlooked in previous work. 
In DILN, the ILEM block first constructs and encodes the interest life-cycle features based on their recent activity patterns under the candidate interest. 
Subsequently, the ILFM performs personalized transformation on both the input and intermediate layers, leveraging the life-cycle features to enhance the capability to capture intricate patterns between interest life-cycles and diverse recommendation tasks.
Through offline experiments on both public and industrial datasets, we demonstrate the effectiveness of the DILN method. Currently, DILN has been deployed on the Lofter platform, achieving a +0.38\% improvement in CTR, +1.04\% improvement in CVR and +0.25\% improvement in Duration.



\bibliographystyle{ACM-Reference-Format}
\bibliography{main}


\end{document}